\documentclass[reprint,amsmath,amssymb,showpacs]{revtex4-1}
\usepackage{epsfig}
\usepackage{amsmath}
\usepackage{amssymb}
\usepackage{graphicx}
\usepackage{dcolumn}
\usepackage{hyperref}
\usepackage{bm}
\usepackage{CJK}

\begin{document}

\title{Local sheet conductivity and sheet current density mapping using a single scanning voltage probe}
\author{Weigang Wang}%(Íõκ¸Õ)}
\author{Malcolm R. Beasley}
\email{beasley@stanford.edu}
\affiliation{Geballe Laboratory for Advanced Materials,\\
Stanford University, Stanford, CA 94305
}%
\date{\today}

\begin{abstract}
 We demonstrate how a single scanning voltage probe can be used to map the local conductivity and current density in a thin film with no a priori knowledge of the geometry of the electrical contacts.  With state-of-the-art scanning voltage probes, under appropriate conditions such mapping should be possible down to nanometer scales.   The technique requires two non-colinear voltage scans.  When only one voltage map is available, determination of the conductivity is not possible because the solution to the governing equation is not unique.  The only restriction on the technique is that the sheet conductivity is a local function of position.
\end{abstract}

\maketitle

Determination of the local variations in the properties of materials is important in both science and technology.  For example, these variations  are important in understanding the spatially averaged macroscopic properties of a material and, in particular, in understanding the microscopic processes from which the macroscopic  properties emerge.  They can also reveal new physical effects. For these reasons characterization of the of size and nature of inhomogeneities in a material is an important  part of materials physics.\cite{Waseda1993,Karppinen1999,Mutin2009,Duarte2012}

%\href{http://www.ingentaconnect.com/content/els/0927796x/1999/00000026/00000003/art00006}{Ref2,}\href{http://pubs.acs.org/doi/abs/10.1021/cm802348c}{Ref3,}\href{http://www.sciencedirect.com/science/article/pii/S0300944011003481}{Ref4}

The revolution in local scanning probes has greatly enhanced the ability to locally characterize materials in all sorts of ways.  However, only recently have reliable means of measuring the electrical transport potential of thin films become practical on the nanoscale, and these advances have opened up  the study of transport processes on these very short  length scales \cite{Briner1996,Rozler2008,Homoth2009,Ji2012}.  However, determination of the local sheet conductivity $\sigma(x,y)$ using these potential probes has proved intractable due to the presumed need to know indepedently the local sheet current density.

In this {\it Letter} we show that provided that the relationship between the sheet current density $\vec j(x,y)$ and the electric field $\vec E(x,y)$ (defined as the gradient of the electro-chemical potential) is local (i.e.,  $\vec j(x,y)=\sigma(x,y)\vec E(x,y)$), a single scanning voltage probe can be used to infer spatial variation of the local conductivity  with no a priori knowledge of the current distribution.  The spatial resolution of the conductivity map is limited by the spatial resolution of the potential measurement and can be on the nanoscale using the highest resolution scanning potentiometers.  The possible applications of this technique range from the practical (e.g., determining the local doping of a semiconductor \cite{French2009}) to the exotic (e.g., identifying very high temperature  trace superconductivity, which is important in the search for higher temperature superconductors).
%\href{http://www.annualreviews.org/doi/abs/10.1146/annurev-matsci-082908-145350}{Ref9}

Existing methods of mapping local conductivity include: (1) micro four-point measurements,  using either four independent scanning probes\cite{Shiraki2001,Ishikawa2005,Guise2005,Higuchi2011,Nakayama2012} or monolithic four-point probes in which all four probes are fabricated on a common substrate that is then scanned as a single unit\cite{Petersen2002a,Petersen2002,Ju2005,Gammelgaard2008,Ansbaek2009}; and (2) scanning microwave impedance microscopy \cite{Lai2008,Kundihikanjana2011,Lai2011,Tselev2012,Chisum2012}, in which the local complex impedance of the sample is measured and from which the local  conductivity can be determined. Both approaches in effect measure the average conductivity over the size of the probe.  The micro four-point probe methods typically have sub-micron spatial resolution and the determination of the conductivity is straightforward.  The most advanced scanning microwave impedance probes have 50nm resolution\cite{Tselev2012} but require simulation/calculation in order to convert the measured impedance signal to the local conductivity\cite{Torigoe2012}. All of these approaches involve a surface measurement and therefore are most appropriate for measuring thin films or surface properties.

More specifically, the approach proposed here is a four-point potentiometric measurement in which three of the probes are fixed macroscopic contacts, and the fourth is either a scanning tunneling potentiometer tip \cite{Muralt1986,Kim2007,Rozler2008,Druga2010,Ji2012} or a scanning conducting atomic force microscope tip \cite{Hersam1998,Nakamura2005,Miyato2010,Ohashi2010}.  The approach is insensitive to the exact shape and/or location of the fixed contacts.  It also permits determination of the local sheet current density as a function of position as well as the conductivity. If the transport is nonlocal, then it applies only on length scales larger than the kernel that describes the nonlocality.  In particular, at short enough length scales (e.g., on atomic length scales) the transport will surely become nonlocal and even the meaning of the measured transport potential (here taken as the electro-chemical potential) becomes a complex issue\cite{Wang2010}.  Establishing experimentally at what length scale our formalism breaks down would be an interesting physical result. %Reference WW and MRB on the Archive.
\begin{figure}
\begin{center}
\includegraphics[width=3in,bb=0 0 400 320]{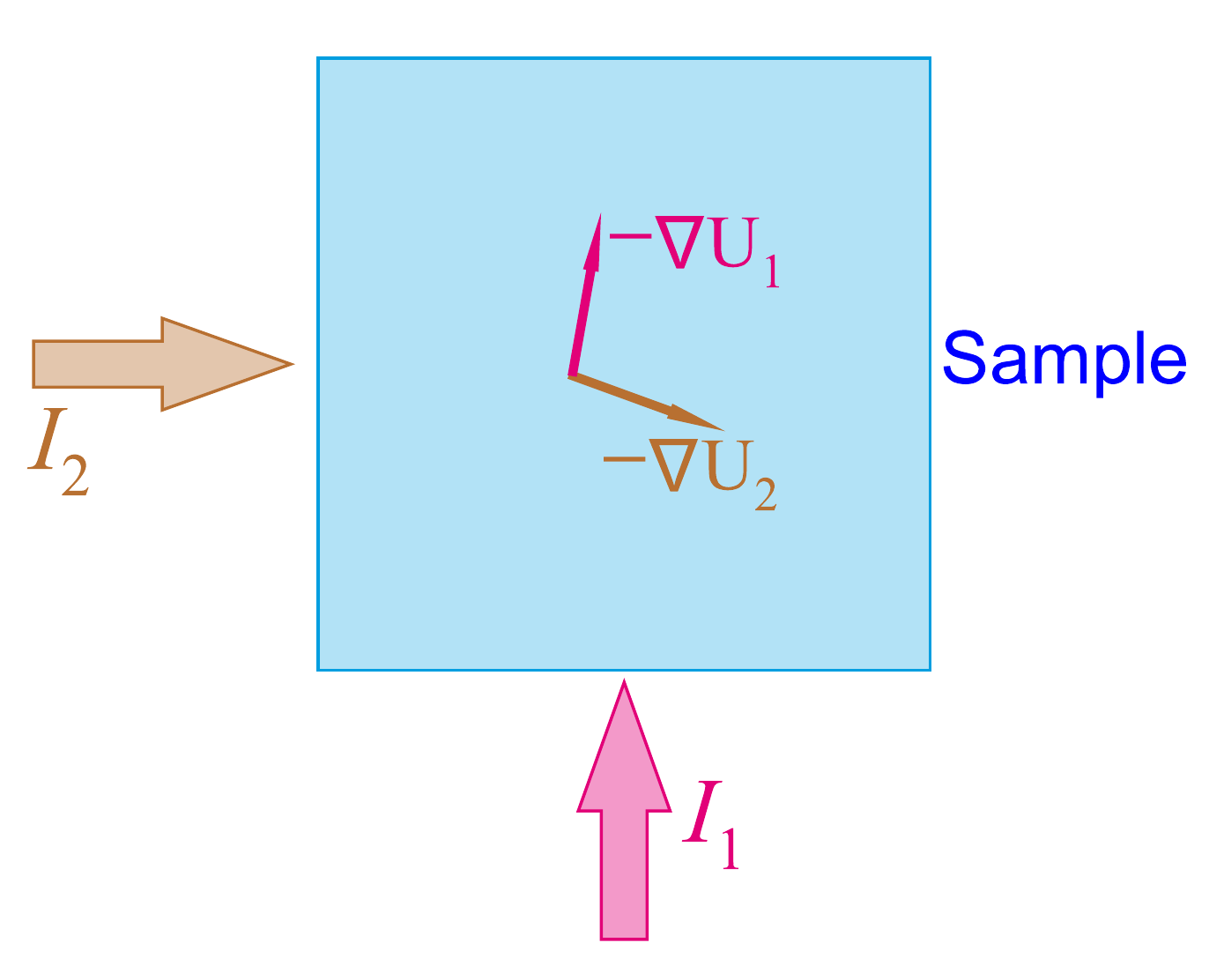}
\caption{We propose to use local voltage probes to measure two voltage scans obtained with currents at different directions. At a specific point, the local electric fields are not parallel, and $\nabla(\ln\sigma)$ is fixed at each position from equation (\ref{onemap}), hence the solution of $\sigma$ is unique.}\label{TwoScans}
\end{center}
\end{figure}

\begin{figure*}
\begin{center}
\includegraphics[width=5.5in,bb=0 0 1000 1431]{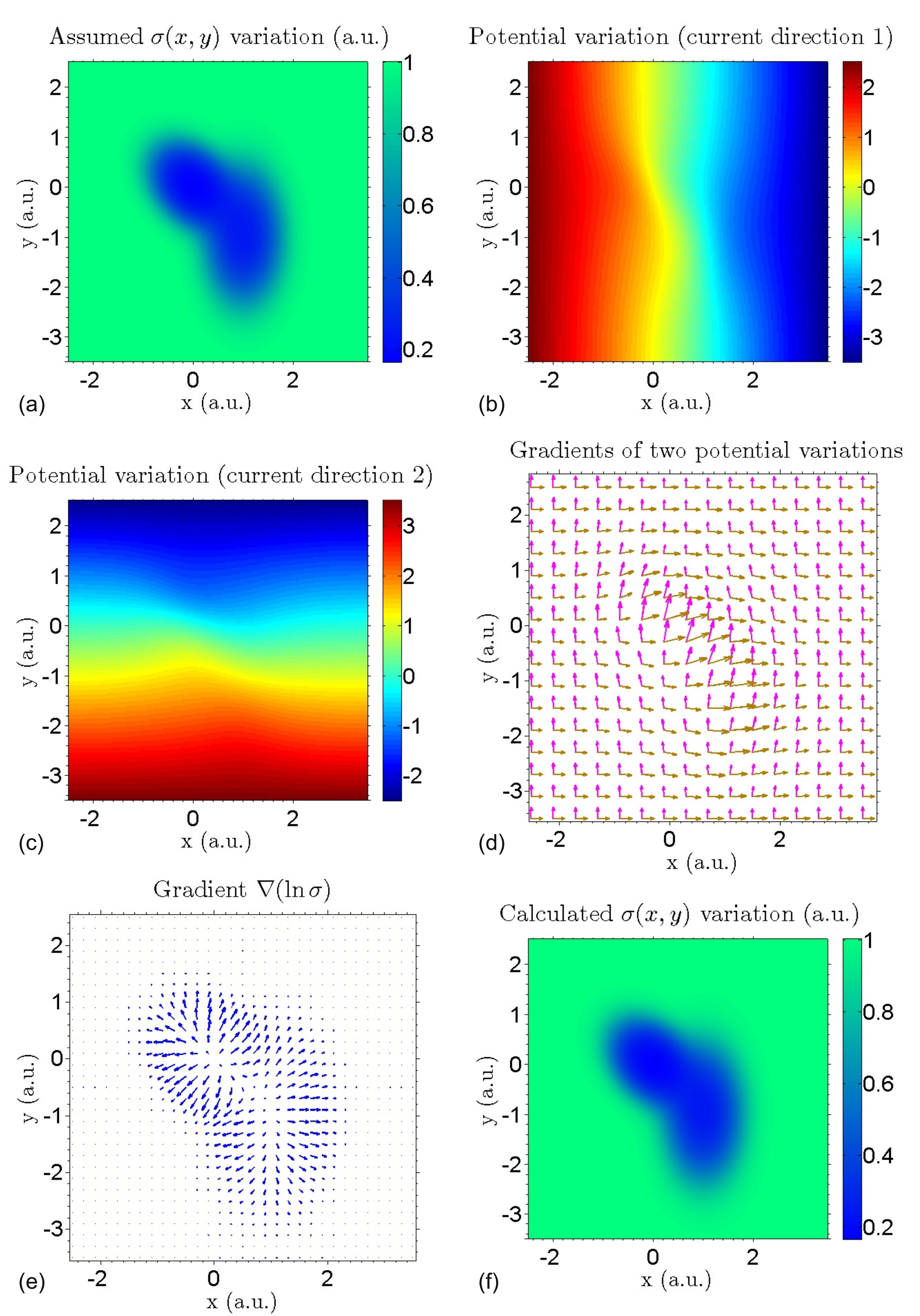}
\caption{Local conductivity mapping from simulated voltage variations. (a) Assumed spatial variation $\sigma(x,y)$ according to equation (\ref{simu1}). (b) Simulated voltage variation from equation (\ref{master}) with current largely in $x$-direction; and (c) simulated voltage variation with current largely in $y$-direction. (d) Calculated local electric fields from simulated voltage variations (b) and (c), sampling is more sparse than actual calculated data density. (e) Calculated gradient field $\nabla(\ln\sigma)$ from equation (\ref{twoscan3}), sampling is more sparse than actual calculated data density; and (f) calculated $\sigma(x,y)$ from the gradient field in (e), according to equation (\ref{twoscan4}).}\label{simulation}
\end{center}
\end{figure*}

The equation governing  local transport is:
\begin{equation}\label{master}
\nabla\cdot\vec j=-\nabla\cdot(\sigma\nabla U)=0\ ,
\end{equation}
where $U$ is the local transport potential, and $\vec j,\sigma$ and $U$ are all assumed henceforth to be functions of position.

The most common application of equation (\ref{master}) is to calculate the local variation of $U(x,y)$ for a given $\sigma(x,y)$ in the presence of a transport current applied through current contacts of known geometry.  Such calculations are straightforward and well defined. On the other hand, in our case, it is only the local variation of $U(x,y)$ that is known, and one wants to determine  $\sigma(x,y)$ and $\vec j(x,y)$.  If only a single measurement of $U(x,y)$ is available,  there is no unique solution for $\sigma(x,y)$ unless other assumptions are made.\footnote{In a recent publication\cite {Li2013}, an additional assumption that the curl of current is zero is used in order to calculate the conductivity from one voltage map. In our opinion, this constraint on the current distribution can be lifted. The reason is that, as we mentioned earlier, in the physical problem with known conductivity variation and a current through the sample, and a voltage variation is to be solved, such constraint on the sheet current density is not necessary; hence when inverting the problem to one of obtaining conductivity from voltage maps in our case, this constraint is also not necessary. Note also that in the special case described by equations (\ref {special1}) through (\ref{special3}), the curl of the sheet current density is not zero.}  However, as we show below, if $U(x,y)$ is known for two (non parallel) configurations of the applied current, a unique solution for $\sigma(x,y)$ can be determined, and this is the basis of the method for local conductivity mapping being presented in this $\it{Letter}$.

Let us now turn to a more formal discussion of these issues.   First rewrite equation (\ref{master}) in the form:
\begin{equation}\label{onemap}
\nabla(\ln\sigma)\cdot\nabla U_0=-\nabla^2U_0\ .
\end{equation}
Here $U_0$ is the local transport voltage measured for a single configuration of the applied current. The reason why the above equation does not have a unique solution for $\sigma$ is now clear.  The equation only determines the projection of gradient $\nabla(\ln\sigma)$ onto the local electric field $\nabla U_0$.  There is no information about the variation of $\sigma$ in the direction perpendicular to $\nabla U_0$.  Stated more formally, the case above leads to the situation that  if $\sigma_1$ is a solution to equation (\ref{onemap}), solution $\sigma_2$ can also be a solution as long as:
\begin{equation}\label{onemap1}
\nabla\left(\ln\left(\frac{\sigma_2}{\sigma_1}\right)\right)\cdot\nabla U_0=0\ .
\end{equation}

A simple example of the previous situation is the following. Suppose the measured local voltage variation is in the form:
\begin{equation}\label{special1}
U_0(x,y)=-Ex\ ,
\end{equation}
where $E$ is the electric field on the sample due to the current through the sample. In this case the measured $U_0$ is a slanted plane in $(x,y)$ space. Does this mean the conductivity is homogeneous in the sample? The answer is no, because one solution to equation (\ref{master}) is:
\begin{equation}\label{special2}
\sigma(x,y)=f(y)\ ,
\end{equation}
\begin{equation}\label{special3}
j_x=f(y)E\ ,
\end{equation}
where $f$ is any function of coordinate $y$. From this example, it can be seen that while the conductivity variation along the electric field lines is fixed by the measured voltage (equation (\ref{onemap})), this variation is free when crossing different electric field lines, hence a non-unique solution for $\sigma$.

In equation (\ref{onemap}), the solution can be uniquely fixed if $\sigma$ is known along a curve not parallel to the electric field. However, this is already an assumption on the variation of $\sigma$, which ideally should be avoided because $\sigma$ is the variable to be solved.

The capability of obtaining a conductivity map from voltage scans without any assumptions about  $\sigma$ can be realized if at least two scans, $U_1$ and $U_2$, are performed with currents in different directions through the sample (see Fig. \ref{TwoScans}). Note that in equation (\ref{onemap}), the gradient $\nabla(\ln\sigma)$ is only free in one direction (perpendicular to the local electric field), and that if another voltage variation is measured, with a different current direction, such that the projection of this gradient onto another direction is also obtained, the gradient will be entirely fixed, hence a unique solution of $\sigma(x,y)$ will be possible.

Let us calculate the gradient $\nabla(\ln\sigma)$ explicitly. From the two voltage scans $U_1$ and $U_2$, one obtains:
\begin{equation}\label{twoscan1}
\nabla(\ln\sigma)\cdot\nabla U_1=-\nabla^2U_1\ ,
\end{equation}
\begin{equation}\label{twoscan2}
\nabla(\ln\sigma)\cdot\nabla U_2=-\nabla^2U_2\ .
\end{equation}
Here, $\sigma(x,y)$ is a property of the sample and therefore independent of the direction of the current.  Thus,  denoting $-\nabla U_{1,2}$ as $\vec E_{1,2}$, one gets:
\begin{equation}\label{twoscan3}
\left(\begin{matrix}
\nabla(\ln\sigma)|_x\\
\\
\nabla(\ln\sigma)|_y
\end{matrix}\right)
=\left(
\begin{matrix}E_{1x}&E_{1y}\\
\\E_{2x}&E_{2y}
\end{matrix}\right)^{-1}\left(\begin{matrix}\nabla^2U_1\\
\\\nabla^2U_2\end{matrix}\right)\ .
\end{equation}

From the above equation, the gradient field $\nabla(\ln\sigma)$ can be calculated at each location from the measured voltage maps, and hence $\sigma$ can be calculated from:
\begin{equation}\label{twoscan4}
\sigma(x,y)=\sigma(x_0,y_0)\exp\left(\int_{x_0,y_0}^{x,y}\nabla(\ln\sigma)\cdot \vec dl\right)\ ,
\end{equation}
which determines $\sigma(x,y)$ up to an arbitrary scale factor $\sigma(x_0,y_0)$. This scale factor can be obtained from other measurements, e.g., spatially averaged macroscopic transport measurement. Given $\sigma(x,y)$ and $\vec E(x,y)$, the local current density $\vec j(x,y)$ can also be determined.

%A critical issue in this approach is that $\sigma(x,y)$ is assumed to be the same for the two different current distributions. It is possible that due to local heating, different current distributions lead to different $\sigma$ variations, especially given that to obtain high signal to noise ratio, a large current density is usually applied through the sample for high resolution voltage probing. This possible issue can be solved by applying both $I_1$ and $I_2$ simultaneously, e.g., with the two currents separated in the frequency domain. With this configuration and simultaneous local voltage measurements at the same point corresponding to the two currents, it can be guaranteed that the local $\sigma(x,y)$ is the same for the two current distributions.

For pedagogical purposes and to show how even the simplest algorithm can be used to analyze potential maps, we now present a numerical example of the above approach. The local conductivity variation was arbitrarily chosen to have a non-trivial shape for these demonstration purposes, specifically:

\begin{widetext}
\begin{equation}\label{simu1}
\sigma(x,y)=\sigma_0\left(1+5\exp\left(-(x-y)^2-2(x+y)^2\right)+3\exp\left(-3(x-1)^2-(y+1)^2\right)\right)^{-1}\ ,
\end{equation}
\end{widetext}
and two simulations were carried out for voltage drops largely in $x$ and $y$ directions, respectively (Fig. \ref{simulation}(b) and (c)). In order to provide simulated potential data, we calculated $U$ from the given $\sigma(x,y)$, the boundary conditions are constant voltages along left/right boundaries for Fig. \ref{simulation}(b), and constant voltages along top/bottom boundaries for Fig. \ref{simulation}(c). The other two boundaries are assumed to have Neumann type conditions $\hat n\cdot\nabla U=0$ for convenience. Note, however, that our procedure for obtaining $\sigma(x,y)$ from the two voltage maps does not depend on boundary conditions (i.e., no information from the boundary conditions is required).

From the calculated local electric field distribution (Fig. \ref{simulation}(d)), one can see that the local conductivity variation causes significant disturbance to the voltage variation, nonetheless, the electric fields at each point are not parallel, hence equation (\ref{twoscan3}) can be used to calculate the gradient field $\nabla(\ln\sigma)$ from these electric field vectors and $\nabla^2U_{1,2}$, which are also numerically calculated from both voltage variations (see Fig. \ref{simulation}(e)).

Finally, from the gradient field $\nabla(\ln\sigma)$, equation (\ref{twoscan4}) was used to calculate the variation $\sigma(x,y)$, plotted in Fig. \ref{simulation}(f). Comparison between Fig. \ref{simulation}(a) and (f) immediately reveals the success in recovering $\sigma(x,y)$ from the two voltage maps. Again, note that only the two voltage maps were used to calculate $\sigma(x,y)$, without any assumptions on the variation of $\sigma(x,y)$ itself. Because the conductivity variation is determined up to an arbitrary scale, in plotting Fig. \ref{simulation}(f) we have scaled the calculated result to the assumption in (a) for convenience of comparison.

%The above demonstration is intended to show that the local conductivity is fixed when two voltage maps are taken. The procedure of obtaining such conductivity mapping, however, may not be optimized from the perspective of computational physics. The success of recovering to the original assumed conductivity map is in part due to the fact that no noise was assumed in the voltage maps. In reality, noise when obtaining the voltage maps, errors in the instrument displacement (i.e. errors in $x$ and $y$), non-local current-electric field relation, and anisotropy of the material are all issues to consider. These issues are to be pursued in a separate paper. We would like to point out that, when the issues of noise and errors in $x$ and $y$ are properly addressed, the exercise of obtaining $\sigma(x,y)$ from measured voltage maps can be a quantitative test of either the anisotropy of the material or the locality of the current-electric field relation. One can thus answer, quantitatively, the question of whether or not the local transport in the material obeys equation (\ref{master}), i.e., whether or not the transport is classical diffusive.

Note that the methodology presented here for determining the local sheet conductivity is not optimized for considerations such as discontinuities in the local conductivity (e.g., a grain boundary) or anisotropy in the material properties. These will require a more elaborate theoretical treatment.  Also, we have not considered the practical matter of the effect of noise in the measurement.  Finally, as already noted, the breakdown of our methodology at very short length scales would provide a direct measurement of the onset of non-locality in the transport.

In conclusion, we have demonstrated that by using two voltage maps with currents in directions at an angle, variations in the local sheet conductivity and local distribution of the sheet current density can be mapped without any a priori assumptions about the conductivity itself.

\begin{acknowledgements}
We would like to thank An-Ping Li for discussions. This work was supported by the Air Force Office of Scientific Research MURI Contract \# FA9550-09-1-0583-P00006. One of us (W.W.) further acknowledges the generous support of a Stanford Graduate Fellowship.
\end{acknowledgements}

%\bibliography{WGWThesis}


\end{document}